\documentclass[10pt,twoside]{article}
\usepackage{amsmath}
\usepackage{amsfonts}
\usepackage{amssymb}
\usepackage{color}
\usepackage{graphicx}
\usepackage[all]{xy}
\usepackage{color}
\usepackage[T1]{fontenc}

\usepackage{graphicx}

\begin{document}

\title{Polytropic equation of state and primordial quantum fluctuations}
\author{R. C. Freitas\thanks{%
e-mail: \texttt{rodolfo.camargo@pq.cnpq.br}} and S.V.B. Gon\c{c}alves\thanks{%
e-mail: \texttt{sergio.vitorino@pq.cnpq.br}} \\
\\
\mbox{\small Universidade Federal do Esp\'{\i}rito Santo,
Departamento
de F\'{\i}sica}\\
\mbox{\small Av. Fernando Ferrari, 514 - Campus de Goiabeiras, CEP
29073-910, Vit\'oria, Esp\'{\i}rito Santo, Brazil}}
\date{}
\maketitle

\begin{abstract}
We study the primordial Universe in a cosmological model where inflation is driven by a fluid with a polytropic equation of state $p = \alpha\rho + k\rho^{1 + 1/n}$. We calculate the dynamics of the scalar factor and build a Universe with constant density at the origin. We also find the equivalent scalar field that could create such equation of state and calculate the corresponding slow-roll parameters. We calculate the scalar perturbations, the scalar power spectrum and the spectral index. \vspace{0.7cm}
\end{abstract}

PACS number(s): 98.80.-k, 98.80.Cq, 04.30.-w \vspace{0.7cm}














\section{Introduction}

Recently the Bose-Einstein condensate (BEC) dark matter (DM), a cosmological model based on condensate state physics, has appeared in several works as an attempt to explain the origin and nature of DM \cite{bec3, harko1, bec1, bec2, harko2, freitas.bec, chavanis1} and was initially used to describe DM halos. The equation of state of the BEC can be found from the Gross-Pitaevskii (GP) equation \cite{bec3, harko1} and is given by
\begin{equation}
\label{BECeos}
   p = \frac{2\pi \hbar^2l_s}{m^3} \rho^2 \quad ,
\end{equation}
where $l_s$ is the scattering length, $\hbar = h/2\pi$ is Planck's constant, $\rho$ is the density distribution of the single component BEC DM and $m$ is the mass of particles that have been condensed.

Assuming the hypothesis that the cold dark matter in a galaxy is in form of BEC the density distribution of the static gravitationally bounded single component BEC DM is given by $\rho(r)=\rho_{\ast}\sin{kr}/kr$, where $\rho_{\ast}=\rho(0)$ is the density in the center of the condensate and $k$ is a constant. Giving the conditions $\rho({\cal R})=0$ and $k{\cal R}=\pi$, where ${\cal R}$ is the condensate radius, the condensate DM halo radius can be fixed as ${\cal R}=\pi(\hbar^{2}l_{s}/ G m^{3})^{1/2}$. The calculated total mass of the condensate DM halo is $M=4\pi^{2}(\hbar^{2}l_s/Gm^{3})^{3/2}\rho_{\ast}=4{\cal R}^{3}\rho_{\ast}/\pi$. So the mass of the particles of the condensate is \cite{harko1}
\begin{equation}
   m=\left(\frac{\pi^2\hbar^2l_s}{G{\cal R}^2}\right)^{1/3} \approx 6.73\times10^{-2}\left(\frac{l_s}{1~\textrm{fm}}\right)^{1/2}\left(\frac{{\cal R}}{1~\textrm{kpc}}\right)^{-2/3}~\textrm{e V} \quad.
\end{equation}

The Bose-Einstein condensation process, that is a very well observed phenomenon in terrestrial experiments, occurs when a gas of bosons is cooled at very low temperatures, near absolute zero, what makes a large fraction of the particles occupy the same ground state. The BEC model can also be applied to cosmology in order to describe the evolution of the recent Universe. In this attempts it can be assumed that this kind of condensation could have occurred at some moment during the cosmic history of the Universe. The cosmic BEC mechanism was broadly discussed in \cite{bec1, bec2}. In general the BEC takes place when the gas temperature is below the critical temperature $T_{\textrm{crt}}<2\pi\hbar^{2}n^{2/3}/mk_{B}$, where $n$ is the particles density, $m$ is the particle mass and $k_B$ is the Boltzmann's constant. Since in an adiabatic process a matter dominated Universe behaves as $\rho\propto T^{3/2}$ the cosmic dynamics has the same temperature dependence. Hence we will have the critical temperature at present $T_{\textrm{crt}}=0.0027~\textrm{K}$ if the boson temperature was equal to the radiation temperature at the red-shift $z=1000$. During the cosmic adiabatic evolution the ratio of the photon temperature and the matter temperature evolves as $T_{\textrm{r}}/T_{\textrm{m}}~\propto~a$, where $a$ is the scale factor of the Universe. Using as value for the present energy density of the Universe $\rho=9.44\times10^{-30} \textrm{g}/\textrm{cm}^{3}$ BEC will happens if the boson mass satisfies $m<~1.87~\textrm{eV}$.

Recently the cosmological process of the condensation of DM was investigated \cite{harko2, freitas.bec} and in this model it is assumed that the condensation process is a phase transition that occurs at some time during the history of the Universe. In this case the normal bosonic DM cools below the critical condensation temperature, that turns to be memorable to form the condensate in which all particles occupy the same ground state. In this new period the two phases coexist for some time until all ordinary DM is converted into the condensed form, when the transition ends. The time evolution of cosmological parameters as the energy density, temperature, scale factor and both scalar and tensorial perturbations is changed during the phase transition process.

There are other cosmological scenarios with BEC. In \cite{harko3}, the author consider a Post-Newtonian cosmological approach to study the global cosmological evolution of gravitationally self-bound BEC dark matter and the evolution of the small cosmological perturbations.  The author verifies that the presence of BEC dark matter changes the cosmological dynamics of the Universe. In \cite{chavanis1} the author do the same as in \cite{harko3} with the generalized EoS $p = (\alpha\rho + k\rho^2)~c^2$. Optimal parameters in good agreement with the $\Lambda$CDM model are found. A natural justification and generalization of the adhesion model, the Burges equation and the cosmological Kardar-Parisi-Zhang equation, that describes the large-scale structure of the Universe but it is introduced heuristically, are studied in \cite{chavanis6}.  The possibility that a significant part of the compact astrophysical objects are made of BEC is also considered in \cite{chavanis7}. The power spectrum for the BEC dark matter is studied in \cite{velten}, where the authors limit the mass of the dark matter particle in the range $15~\mbox{meV} < m < 700~\mbox{meV}$. The effects of a finite dark matter temperature on the cosmological evolution of the BEC dark mater systems is verified in \cite{harko4}. In this paper the authors have shown that the presence of thermal excitations leads to an overall increase in the expansion rate of the Universe. With the aim of understanding whether the gravitational interactions of axions can generate entropy, the behavior of the axions during non-linear galaxy evolution is studied in \cite{sacha}, but the assumption that the axion field can form a Bose-Einstein condensate is not yet confirmed. In \cite{shapiro} the authors assume that the dark matter particles are described by a spin-0 scalar field, called scalar field dark matter. These bosonic particles are ultralight, with masses down to the order of $10^{-33}~\mbox{eV/}\mbox{c}^2$. An ultralight phase-space density is suggests and there is the possibility of formation of a Bose-Einstein condensate. They shown that the scalar field dark matter is compatible with observations of the cosmic microwave background and the abundance of the light elements produced by Big Bang nucleosynthesis. With a relativistic version of the Gross-Pitaevski equation, the authors in \cite{bec1, chavanis2, chavanis3} propose i) a novel mechanism of inflation, ii) a natural solution for the cosmic coincidence problem with the transition from dark energy into dark matter, iii) a very early formation of highly non-linear objects like black holes and iv) log-z periodicity in the subsequent BEC collapsing time.

We can generalize the BEC equation o state (EoS) (\ref{BECeos}) \cite{chavanis2, chavanis3} as follows
\begin{equation}
p=\alpha \rho+k\rho^2\quad,
\end{equation}
where $k>0$ represent repulsive and $k<0$ attractive self-interaction and the linear term describe the well known radiation ($\alpha=1/3$), dust matter ($\alpha=0$) and cosmological constant ($\alpha=-1$), and the less known stiff matter ($\alpha=1$). The stiff matter model is a  specific cosmological model where the matter content of the Universe has an equation of state of the form, $p = \alpha\rho$, with $\alpha = 1$, where $\rho$ and $p$ are, respectively, the fluid energy density and pressure \cite{zeldo}. This model can also be described by a massless free scalar field. The energy density of the stiff matter is proportional to $1/a(t)^6$ and this result indicates that there may have existed a phase earlier than that of radiation, where $\alpha = 1/3$ and $\rho\propto 1/a(t)^4$, and after inflation in our Universe, which was dominated by stiff matter. This peculiarity motivated us to investigate their behavior in the analyses made here in this work and to consider the implications of the presence of a stiff matter perfect fluid in FRW cosmological models.

The EoS $p=\alpha \rho+k\rho^2$ is the sum of the linear term and a quadratic term, that describes BECs. At late times, when the density is low, the BECs contribution to the EoS is negligible and the evolution is determined by the linear term. But in the early Universe, when the density is high, the term due to BECs in the EoS is dominant and modifies the dynamics of the Universe. Lately this model was used as a model of the early Universe. We can assume that this EoS holds before radiation era and for the repulsive self-interaction the Universe starts at $t=0$ at a singularity with infinite density but finite radius. For the case of attractive self-interaction the Universe has always existed and for the non-physical limit $t\rightarrow-\infty$ the density tends to a constant value and the radius goes to zero, both exponentially \cite{chavanis2, chavanis3}. 

In this letter we study the generalized EoS \cite{chavanis4}
\begin{equation}
p=\alpha\rho+k\rho^{1+1/n}\quad,
\end{equation}
to describe the physical state of the matter content of the Universe, where $n=1$ and $\alpha=0$ describes cosmological BECs. With the generalized EoS this model can present a phase of early accelerated expansion. It can be also used to describe a phase of late accelerated expansion, depending on the choice of the parameters \cite{chavanis2, chavanis3}. We calculate the primordial cosmological dynamics in this model for $\alpha=1/3$, $\alpha=0$ and $\alpha=1$ \cite{chavanis2}. We also find a scalar potential that can generate this EoS and calculate both scalar and tensorial perturbations \cite{chavanis3}. We study the corresponding slow-roll parameters and the power spectrum and spectral index are calculated.

To motivate the model studied here we can see an analogy between this polytropic equation of state and a cosmological model where the fluid that fills the Universe has an effective bulk viscosity \cite{Barrow}. If we write $p = \alpha\rho - 3H\eta$, where $\eta$ is the viscous coefficient, we have exactly the generalized EoS, when $\eta\propto\rho$ and $H\propto\rho^{1/n}$.

The present letter is organized as follows. In section \ref{sec:GPequation} we introduce the generalized Gross-Pitaevskii equation used to describe the BEC in the short-ranged scale. In section \ref{sec:dinamica} we study the evolution of the Universe filled with fluid described by a polytropic EoS. For the case of a non-singular inflationary Universe we find the scalar potential that could generate the polytropic EoS and calculate the slow-roll parameters. The primordial fluctuations, such as gravitational waves, perturbations for the gravitational potential and for  the density contrast are calculate, and we find the quantum power spectrum and spectral index in section \ref{sec:pertrubacoes}. We present our conclusions and discussions in section \ref{sec:conclusao}.

\section{The generalized Gross-Pitaevskii equation}   
\label{sec:GPequation}                                

 The Gross-Pitaevskii (GP) equation is a long-wavelength theory widely used to describe dilute BEC, but it fails \cite{modbec} in the case of short-ranged repulsive interactions in low dimensions. Therefore the inter-particle interaction term in the GP equation must be modified and in this model the ground state features of the BEC is described by the generalized GP equation \cite{bec3, harko2}
\begin{equation}
   \label{eq.gGP}
   \dot{\imath}\hbar\frac{\partial \phi(t,\vec{r})}{\partial t}=-\frac{\hbar^2}{2m}\nabla^2\phi(t,\vec{r})+mV(\vec{r})\phi(t,\vec{r})+g'(n)\phi(t,\vec{r}) \quad ,
\end{equation}
where $\phi(t,\vec{r})$ is the wave function of the condensate, $m$ is the particles mass, $V$ is the gravitational potential that satisfies the Poisson's equation $\nabla^2V(\vec{r})=4\pi G\rho$, $g'=dg/dn$, $n=|\phi(t,\vec{r})|^2$ is the BEC density and $\rho=mn$. To understand the physical properties of a BEC we can use the Madelung representation of the wave function \cite{bec3, bec1, bec2}, which is
\begin{equation}
   \phi(t,\vec{r})=\sqrt{n(t,\vec{r})}\times e^{\dot{\imath}S(t,\vec{r})/\hbar} \quad ,
\end{equation}
where $S(t,\vec{r})$ has the dimension of an action. This transformation will make the generalized GP equation (\ref{eq.gGP}) breaks into two equations
\begin{eqnarray}
   \frac{\partial \rho}{\partial t} +\nabla \cdot(\rho \vec{v}) & = & 0 \quad , \\
   \rho \left(\frac{\partial \vec{v}}{\partial t}+(\vec{v}\cdot \nabla)\vec{v}\right) & = & -\nabla p\left(\frac{\rho}{m}\right)-\rho\nabla\left(\frac{V}{m}\right) -\nabla V_Q \quad ,
\end{eqnarray}
where $V_Q=-(\hbar^2/2m)\nabla^2\sqrt{\rho}/\rho$ is a quantum potential and $\vec{v}=\nabla S/m$ is the velocity of the quantum fluid. The effective pressure of the condensate \cite{bec1, bec2, harko2} is given by
\begin{equation}
   p\left(\frac{\rho}{m}\right)=g'\rho -g \quad .
\end{equation}
Now writing $g \propto \rho^\gamma$ we find the generalized EoS
\begin{equation}
   \label{eq.eos2}
	  p = k\rho^\gamma \quad,
\end{equation}
where $k$ is a proportionality constant that will be determined in the context of our model, or can be related to the mass and the scattering length of the boson in the case of the long-wavelength theory, and $\gamma\equiv 1+1/n$ is the polytropic index.

\section{Equation of state and cosmic dynamics}
\label{sec:dinamica}                           

Following the present data \cite{wmap1,wmap2,planck1} we assume a flat homogeneous and isotropic Universe, whose geometry is described by the Friedmann-Robertson-Walker metric, given by
\begin{equation}
   \label{eq:metric}
   ds^2=dt^2-a(t)d\vec{x}^2 \quad ,
\end{equation}
where $a(t)$ is the scale factor of the Universe that describes the cosmic evolution, $t$ is the cosmic time and we made the speed of light $c=1$. The gravitational dynamics is given by the Einstein's field equations
\begin{equation}
   \label{eq:einstein}
   R_{\mu\nu}-\frac{1}{2}g_{\mu\nu}R=8\pi GT_{\mu\nu} \quad.
\end{equation}
We also consider the Universe filled by a perfect fluid, described by the energy-momentum tensor
\begin{equation}
   T^{\mu\nu}=(\rho +p)u^\mu u^\nu - pg^{\mu\nu} \quad ,
\end{equation}
where $\rho$ is the density of the fluid, $p$ is the pressure and $g^{\mu\nu}$ is the metric tensor.
 
This perfect fluid has a general equation of state (EoS), presented in \cite{chavanis2,chavanis3}, that is a sum of a standard linear EoS and a polytropic term,
\begin{equation}
   \label{eq:EoS}
	  p=\alpha\rho+k\rho^{1+1/n} \quad,
\end{equation}
where $-1\leq\alpha\leq1$, $k$ is the polytropic constant and $1+1/n$ is the polytropic index. In the linear term $\alpha=-1$ represents vacuum energy, $\alpha=1/3$ is radiation, $\alpha=0$ is  pressureless matter and $\alpha=1$ is stiff matter. The polytropic term may represent self-gravitating Bose-Einstein condensate (BEC) with repulsive ($k>0$) or attractive ($k<0$) self-interaction, where $n=1$ corresponds to the standard BEC.

Here we will consider the high density case, $(1+\alpha+k\rho^{1/n})\geq0$ and $n>0$ to describe the primordial Universe which means that the density decreases with the radius. The case $1+\alpha+\rho^{1/n}\leq0$ represents a phantom Universe, where the density increases with the radius \cite{chavanis4}. In both cases the polytropic term in the EoS (\ref{eq:EoS}) dominates when the density is high and $n>0$ and when the density is low and $n<0$.

For the  EoS (\ref{eq:EoS}) the energy conservation equation is
\begin{equation}
   \label{eq:conservacao}
   \dot{\rho}+3H\rho(1+\alpha+k\rho^{1/n})=0 \quad,
\end{equation}
where overdot denote the derivative with respect to the cosmic time $t$ and $H=\dot{a}/a$ is the Hubble parameter. With $\alpha\neq-1$ this equation is integrated to give
\begin{equation}
\label{eos01}
   \rho = \frac{\rho_{*}}{\left[(a/a_0)^{3(1+\alpha)/n}\mp 1\right]^n} \quad,
\end{equation}
with the minus sign corresponding to $k>0$, the plus sign corresponding to $k<0$, $(a/a_0)=a/a_0$, where $a_0$ is a constant of integration, and $\rho_{*}=\left[\left(1+\alpha\right)/|k|\right]^n$.

For the repulsive self-interaction ($k>0$) the density is defined only for $a_0<a<\infty$, where
\begin{eqnarray}
\frac{\rho}{\rho_{*}}\approx
\begin{cases}
\biggl(\frac{n}{3(1+\alpha)}\bigg)^{n}\frac{1}{((a/a_0)-1)^n} \rightarrow \infty \quad , \quad a\rightarrow a_0 \\
				\quad \quad (a/a_0)^{-3(1+\alpha)} \rightarrow 0 \quad \quad \quad , \quad a\rightarrow \infty
\end{cases}
		  \quad .
\end{eqnarray}

In the case of an attractive self-interaction ($k<0$) the density is defined for $0<a<\infty$, and
\begin{eqnarray}
   \frac{\rho}{\rho_{*}}\approx
	   \begin{cases}
	      \quad \quad \quad 1 \quad \quad \quad , \quad a \rightarrow  0 \\
				(a/a_0)^{-3(1+\alpha)} \rightarrow 0 \quad , \quad a \rightarrow \infty
	   \end{cases}
		  \quad ,
\end{eqnarray}
with, in the same limits, $p=-\rho_{*}$ and $p\rightarrow 0$.

As we are analyzing the primordial Universe we can see that, when $a\rightarrow 0$, the density $\rho\rightarrow\rho_*$, and we can identify the Planck density $\rho_* = \rho_P = c^5/G^2~\hbar\approx 5.16\times 10^{99}~\mbox{g/m$^3$}$, where $\rho_P$ corresponds to a maximum value for the density in the limit $a\rightarrow 0$. Likewise, the constant of integration $a_0$ can be considered a reference point for the transition between the vacuum energy era, when the polytropic component dominates the EoS (\ref{eq:EoS}),  and the era dominated by linear term in (\ref{eq:EoS}). When $a \ll a_0$, the scale factor increases exponentially as
\begin{equation}
\label{expo01}
a(t)\propto l_P~e^{(8\pi/3)^{1/2}t/t_P}\quad,
\end{equation}
where $t_P\approx 5.39\times 10^{-44}$ s is the Planck time and $l_P = c~t_P = (G\hbar/c^3)^{1/2}\approx 1.62\times 10^{-35}$ m is the Planck length. From the mathematical point of view, in this scenario there is no primordial singularity since this Universe exists at any time in the past ($a\rightarrow 0$ and $\rho\rightarrow\rho_P$ for $t\rightarrow -\infty$). It is obvious that in the limit $a\rightarrow 0$ a quantum theory of gravity is required. Even so, exponential solution (\ref{expo01}) provides a semi-classical description of the early Universe (see \cite{chavanis8} and (\ref{NSI}) section for more detail).

We can also write the equation (\ref{eq:EoS}) as
\begin{equation}
   \label{eq:EoS1}
   p=\omega(t) \rho  \quad,
\end{equation}
where the effective EoS parameter $\omega(t)$ is
\begin{equation}
   \label{eq:EoSpar}
   \omega(t) =\alpha \pm (\alpha+1) \left(\frac{\rho}{\rho_{*}}\right)^{1/n} = \alpha \pm (\alpha+1) \left((a/a_0)^{3(1+\alpha)/n}\mp 1\right)^{-1} \quad.
\end{equation}
With equation (\ref{eq:EoS1}) we can calculate the sound speed in the fluid, which is
\begin{equation}
   \label{eq:vsom}
	  c_s^2=\left(\frac{n+1}{n}\right)\omega(t)-\frac{\alpha}{n}\quad.
\end{equation}

Once again we can find the limits for both repulsive and attractive self-interaction. For $k>0$ we have
\begin{eqnarray}
   \omega(t)\approx
	   \begin{cases}
	      \alpha+\frac{n}{3\left((a/a_0)-1\right)} \quad \rightarrow \infty \quad , \quad a\rightarrow a_0 \\
				\alpha + \frac{\alpha+1}{(a/a_0)^{-3(1+\alpha)/n}} \rightarrow \alpha  \quad , \quad a\rightarrow \infty
	   \end{cases}
		  \quad ,
\end{eqnarray}
and for $k<0$
\begin{eqnarray}
   \omega(t)\approx
	   \begin{cases}
	      \quad \quad \quad -1 \quad \quad \quad \quad \quad , \quad a \rightarrow  0 \\
				\alpha-\frac{\alpha+1}{(a/a_0)^{-3(1+\alpha)/n}} \rightarrow \alpha \quad , \quad a \rightarrow \infty
	   \end{cases}
		  \quad .
\end{eqnarray}

\subsection{Non-singular inflationary Universe}
\label{NSI}

We assume that the Universe is filled by the fluid with EoS (\ref{eq:EoS}), with $n>0$ and $k<0$. With the metric (\ref{eq:metric}), the Einstein's field equations (\ref{eq:einstein}) and the equation (\ref{eq:EoS}) we find the Friedmann equation
\begin{equation}
   \label{eq:friedmann}
   \frac{1}{(a/a_0)^2}\left(\frac{d{(a/a_0)}}{dt}\right)^2=\frac{8\pi G}{3}\rho_{*}\left(1+(a/a_0)^{3(1+\alpha)/n}\right)^{-n} \quad .
\end{equation}
For small values of scale factor $a$, i.e., when $a \ll a_0$ we have $(a/a_0)\rightarrow 0$ and we can expand the Friedmann equation (\ref{eq:friedmann}), for $x \equiv (a/a_0)^{3(1+\alpha)/n} \ll 1$, as
\begin{equation}
   \left[1+\frac{n}{2}x+\frac{1}{2}\left(\frac{n}{2}-1\right)\frac{n}{2}x^2+\frac{1}{6}\left(\frac{n}{2}-2\right)\left(\frac{n}{2}-2\right)\frac{n}{2}x^3 + \mathcal{O}(x^4) \right]\frac{d(a/a_0)}{(a/a_0)}=\sqrt{\frac{8\pi G\rho_{*}}{3}}dt \quad.
\end{equation}
We impose the condition $3(1+\alpha)/n\geq 1$, which means that
\begin{equation}
   \label{eq:cond}
	 n\leq 3(1+\alpha) \quad,
\end{equation}
and we keep only the null order terms to find
\begin{equation}
   (a/a_0) \propto e^{t H_{*}} \quad ,
\end{equation}
where $H_{*}=\sqrt{\frac{8\pi G\rho_{*}}{3}}$. This means that under these conditions the Universe is inflationary and the singularity can be found at the non-physical limit $t\rightarrow -\infty$ with a nearly constant finite density. This indicate that the Universe can start at any time $t_*$, which we will define as $t_*=0$.

For the case of $n>0$ and $k<0$ and $a \ll a_0$ the fluid with EoS (\ref{eq:EoS}) behaves like the vacuum energy, with constant density. The value of $\rho_{*}$ defines a maximum value for the density and it can be limited by the value of the Hubble parameter at the end of inflation \cite{planck2}. With $\rho_{*}$ we fix \cite{chavanis2}
\begin{equation}
   k=-\frac{(1+\alpha)}{\rho_*^{1/n}} \quad .
\end{equation}

The Friedmann equation is
\begin{equation}
\label{Frie}
   \frac{\dot{a}}{a}\approx\sqrt{\frac{8\pi G}{3}\rho_{*}} \quad ,
\end{equation}
and we can calculate the scale factor for $a \ll a_0$
\begin{equation}
   \label{eq:escalainflacao}
   a \approx e^{t H} \quad ,
\end{equation}
where $H\approx\sqrt{\frac{8\pi G}{3}\rho_{*}}$.

After the inflationary stage, for $a \gg a_0$, the linear term of the EoS (\ref{eq:EoS}) dominates, which means that at some point the inflationary stage will come to an end, and the density will behave as stiff matter, radiation or dust, depending on the value of $\alpha$. If $\alpha=1/3$, for example,  the Universe undergoes a radiation era, which would correspond to the standard Universe model. Here we are interested only in the inflationary phase, so we will deal only with solution (\ref{eq:escalainflacao}), which is valid for any $\alpha$ and $n$, since condition ($\ref{eq:cond}$) is true and $k<0$, $\alpha\neq-1$ and $n>0$.

In case that the linear term has a proportionality constant with value $\alpha = -1$ we can solve the conservation equation (\ref{eq:conservacao}) in order to find a density $\rho$ that decays with the scale factor $a$ only if $k>0$ and $n>0$, and it can be found \cite{chavanis2} that
\begin{equation}
   \rho = \frac{\rho_*}{\left[\ln{(a/a_0)}\right]^n} \quad ,
\end{equation}
where $\rho_* \equiv (n/3k)^n$. In this case the density $\rho(a)$ is defined for $a \geq a_0$, and we can't reproduce the behavior of an inflationary Universe. We will not study the perturbations for the case $\alpha = -1$.

\subsection{Slow-roll formalism}

Here we will represent our fluid as a scalar-field $\phi$ and we find the scalar potential $V(\phi)$ \cite{muka,inflation} that generates the EoS (\ref{eq:EoS}). The scalar field representation can more conveniently retain the features we could expect from fluids with negative pressure, responsible for the inflationary phase, mainly for those that are interesting for cosmology, as the scenarios resulting from phase transitions \cite{sergio}. The scalar field must obey the Klein-Gordon equation
\begin{equation}
   \label{eq:KG}
   \ddot{\phi}+3H\dot{\phi}+V_{,\phi}=0 \quad,
\end{equation}
where $V_{,\phi}=dV/d\phi$, and we define
\begin{eqnarray}
   \label{eq:rhophi}
   \rho  & = & \frac{\dot{\phi}^2}{2}+V(\phi) \quad , \\
	 \label{eq:rhop}
	 p & = & \frac{\dot{\phi}^2}{2}-V(\phi) \quad .
\end{eqnarray}
Inflation will only occur \cite{muka,inflation} if
\begin{eqnarray}
   \frac{\dot{\phi}^2}{2} & \ll & V \propto H^2 \quad , \\ \nonumber
	 \label{eq:condinflacao}
	 |\ddot{\phi}| & \ll & 3H\dot{\phi} \approx |V_{,\phi}| \quad ,\\
	 V_{,\phi \phi} & \ll & V \quad . \nonumber
\end{eqnarray}
We combine equations (\ref{eq:rhophi}) and (\ref{eq:rhop}) to find
\begin{eqnarray}
   \label{eq:phi}
   \dot{\phi}^2 & = & (\omega(t)+1)\rho  \quad, \\
	 \label{eq:V}
	 V(\phi) & = & \frac{\rho }{2}(1-\omega(t)) \quad.
\end{eqnarray}

\begin{figure}[ht]
   \centering
   \includegraphics[width=0.8\textwidth]{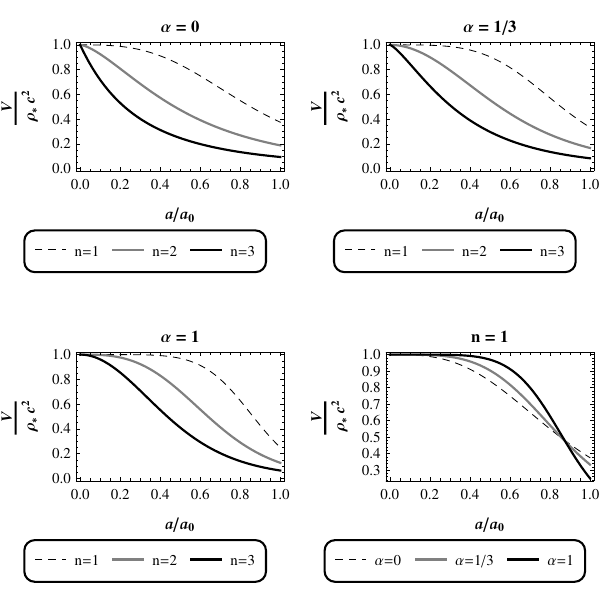}
   \caption{The potential for the scalar field as a function of the scale factor.}
   \label{fig:potencial}
\end{figure}

In order to find how the scale factor $a$ varies with the scalar field $\phi$ we use the chain rule and combine equation (\ref{eq:phi}) with the Friedmann equation (\ref{Frie}) to have
\begin{equation}
   \frac{d \phi}{d a}=\sqrt{\frac{3}{8\pi G}}\frac{\sqrt{\omega+1}}{a} \quad .
\end{equation}

With help the equation (\ref{eq:EoSpar}) we can invert the above solution to give us the solution
\begin{equation}
   \label{eq:Rphi}
   (a/a_0)^{3(1+\alpha)/n}=\sinh^2(\psi) \quad,
\end{equation}
with $\psi$ defined as
\begin{equation}
   \psi = \sqrt{6\pi G\frac{1+\alpha}{n^2}}\phi \quad .
\end{equation}

With equations (\ref{eq:V}) and (\ref{eq:Rphi}) combined we have \cite{chavanis3}
\begin{eqnarray}
\label{pot01}
   V(\phi) & = & \frac{\rho_* }{2}\left[\frac{1-\alpha}{\left(1+(a/a_0)^{3(1+\alpha)/n}\right)^n}+\frac{1+\alpha}{\left(1+(a/a_0)^{3(1+\alpha)/n}\right)^{n+1}}\right] = \nonumber \\
	& = & \frac{\rho_*}{2}\left[\frac{(1-\alpha)}{(\cosh \psi)^{2n}}+\frac{(1+\alpha)}{\left(\cosh \psi\right)^{2(n+1)}}\right] \quad . 
\end{eqnarray}

The inflationary expansion of the Universe will happen while the $\psi \ll 1$. If we expand the scalar potential (\ref{pot01}) for small values of the scalar field $\psi$ we get
\begin{equation}
   V(\phi) \approx \rho_* \left\{1-\frac{(1+\alpha+n)}{2}\psi^2+\left[\frac{(1+\alpha+2n)}{3}+\frac{n(\alpha+n)}{2}\right]\psi^4 \right\} \quad ,
\end{equation}
which resembles a symmetry breaking scalar field potential.

\begin{figure}[ht]
   \centering
   \includegraphics[width=0.8\textwidth]{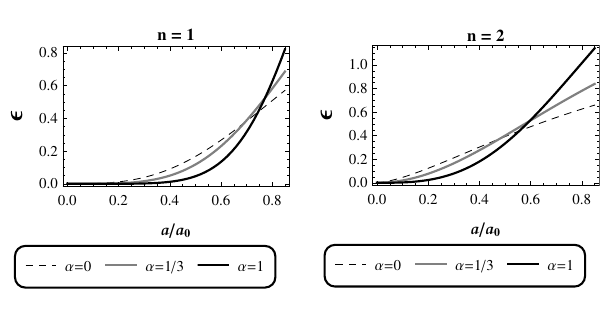}
   \caption{The first slow-roll parameter $\epsilon$ as a function of the scale factor.}
   \label{fig:epsilon}
\end{figure}

The first slow-roll parameter, that is related to the measure of accelerated expansion during inflation, is
\begin{equation}
   \epsilon(t)\equiv -\frac{\dot{H}}{H^2} \quad .
\end{equation}
The accelerated expansion occurs while $\epsilon<1$, and in our model, we have $\epsilon \ll 1$ for $a \ll a_0$. Inflation ends when $\epsilon \approx 1$, and from the evolution of the slow-roll parameter $\epsilon$ in Figure \ref{fig:epsilon} we can clearly see that the inflationary regime will end as the evolution approaches $a_0$.

With the background equations, the EoS parameter (\ref{eq:EoSpar}) and the Klein-Gordon equation (\ref{eq:KG}) we can show that
\begin{equation}
   \epsilon(t)=\frac{3}{2}(1+\omega(t))=4\pi G\frac{\dot{\phi}^2}{H^2} \approx \frac{1}{16 \pi G}\left(\frac{V_{,\phi}}{V}\right)^2 \quad .
\label{slow}
\end{equation}

See Figure \ref{fig:epsilon} for the behavior of the equation (\ref{slow}) as a function of the scale factor $(a/a_0)$ in two different situations, with $n = 1$ and $n = 2$.

The second slow-roll parameter tells us how long the accelerated expansion will be sustained. It is related to the smallness of the second time derivative of the scalar field, and we can write
\begin{equation}
   \eta(t) \equiv -\frac{\ddot{\phi}}{H\dot{\phi}} = \epsilon -\frac{1}{2\epsilon}\frac{d\epsilon}{dN} \approx \frac{1}{8\pi G}\frac{V_{,\phi \phi}}{V} \quad,
\end{equation}
where $dN=Hdt$ and $V_{,\phi \phi}=d^2V/d\phi^2$. During inflation we have that $|\eta|<1$. The slow-roll conditions are
\begin{equation}
   \epsilon,|\eta| \ll 1 \quad.
\end{equation}
See \cite{dodelson} for more discussions on the slow roll parameters.

As already mentioned before, after the inflationary stage, for $a \gg a_0$, the linear term of the EoS (\ref{eq:EoS}) dominates, which means inflation will come to an end, and the Universe will be dominated by the linear component of the EoS (\ref{eq:EoS}). If $\alpha=1/3$, for example, the Universe undergoes a radiation era, which would correspond to the standard Big Bang model, in which the Universe undergoes a radiation dominated era. Although it is possible to reheat the Universe using the scalar-field formalism and the potential (\ref{pot01}), we can not say the same about the fluid formalism. A complete reheating analysis should be made in this model in order to see if the Universe can be reheated.

\section{Primordial quantum perturbations: scalar perturbations}
\label{sec:pertrubacoes}                  

In this section we calculate scalar perturbations generated in the early Universe. Scalar quantum fluctuations can be the source of the seeds that originated the large scale structures we see today. First we introduce the conformal time $\tau$, such that
\begin{equation}
   dt=a(t)d\tau \quad .
\end{equation}
During the inflation we have
\begin{equation}
\tau\equiv\int_{a_e}^a\frac{da}{Ha^2}\quad,
\end{equation}
where $a_e$ is the scale factor at the end of inflation. As $H$ is approximately constant during this period we can consider that
\begin{equation}
\tau\simeq\frac{1}{H}\int_{a_e}^a\frac{da}{a^2}\quad.
\end{equation}
The scale factor at the end of inflation is much large than in the middle, $(a_e \gg a)$. So, we find that
\begin{equation}
   \label{eq:escala}
   a\simeq-\frac{1}{H\tau}\quad .
\end{equation}

In order to find the fluctuations that originated the large scale structures we introduce the perturbed metric
\begin{equation}
   \label{eq:pertscalar}
   ds^2=a(\tau)^2\left[(1+2\Phi)d\tau^2-(1-2\Phi)d\vec{x}^2\right] \quad ,
\end{equation}
where $\Phi(\tau,\vec{x})$ is the gauge-invariant Bardeen's potential  \cite{muka}. We substitute the metric (\ref{eq:pertscalar}) in the Einstein's field equations (\ref{eq:einstein}), and keeping only the first order terms we find
\begin{eqnarray}
   \label{eq:movpertscalara1}
   \nabla^2\Phi-3\mathbb{H}\left(\Phi'+\mathbb{H}\Phi\right) & = & 4\pi Ga^2\delta T^{0}_{~0} \quad , \\
	 \label{eq:movpertscalarb1}
	 \left(\Phi'+\mathbb{H}\Phi\right)_{,i} & = & 4\pi Ga^2\delta T^{0}_{~i} \quad , \\
	 \label{eq:movpertscalarc1}
   \left[\Phi'' + 3\mathbb{H}\Phi'+\left(2\mathbb{H}'+\mathbb{H}^2\right)\Phi\right]\delta^{i}_{~j} & = & -4\pi G a^2\delta T^{i}_{~j} \quad , 
\end{eqnarray}
where $\delta T^{\mu}_{~\nu}$ is the gauge-invariant perturbed stress-energy tensor and $\mathbb{H} = a'/a$ is the Hubble parameter in terms of the conformal time.

Using the hydrodynamics description of the polytropic fluid we first perturb the density $\rho\rightarrow\rho+\delta\rho$ and equations (\ref{eq:movpertscalara1}) and (\ref{eq:movpertscalarc1}) will be
\begin{eqnarray}
   \label{eq:movpertscalara3}
   \nabla^2\Phi-3\mathbb{H}\left(\Phi'+\mathbb{H}\Phi\right) & = & 4\pi a^2G\delta \rho \quad , \\
	 \label{eq:movpertscalarc3}
   \Phi'' + 3\mathbb{H}\Phi'+\left(2\mathbb{H}'+\mathbb{H}^2\right)\Phi & = & 4\pi a^2G\delta p \quad . 
\end{eqnarray}
It is easy to show that $\delta p = c_s^2\delta \rho$. We join equations (\ref{eq:movpertscalara3}) and (\ref{eq:movpertscalarc3}) to find
\begin{equation}    
   \Phi''+3\mathbb{H}\left(1+c_s^2\right)\Phi'+\left[2\mathbb{H}'+\mathbb{H}^2\left(1+3c_s^2\right)+c_s^2k^2\right]\Phi=0 \quad ,
\end{equation}
where $k$ is the modulus of the wave-number and we made $\nabla^2\Phi=-k^2\Phi$. To obtain $\Phi$ during inflation we use the scale factor (\ref{eq:escala}) and the transformations
\begin{eqnarray}
   \Phi & = & a^{\beta}\mu \quad , \\ 
	 \beta=-\frac{3}{2}\left(1+c_s^2\right) & = & \frac{3}{2}\left(\frac{1+\alpha}{n}\right) \quad ,
\end{eqnarray}
and we find
\begin{equation}
   \label{eq:Phi}
   \mu''+\mu\left[\left(k\tau c_s\right)^2-\left(\beta^2+\beta\right)\right]=0 \quad .
\end{equation}
With help of $\mu=\sqrt{|\tau|}g$ we will get the Bessel differential equation
\begin{equation}
   z^2\frac{d^2g}{dz^2}+z\frac{dg}{dz}+\left[z^2-\left(\beta^2+\beta+1/4\right)\right]g=0 \quad ,
\end{equation}
with $z=k|\tau|c_s$. The solution is
\begin{equation}
   \label{eq:solBessel}
   g=C_1\mathcal{H}_\nu^{(1)}(z)+C_2\mathcal{H}_\nu^{(2)}(z) \quad ,
\end{equation}
where $\mathcal{H}_{\nu}^{(i)}(z)$ are the Hankel functions of order $\nu=\sqrt{\beta^2+\beta+1/4}$.

The same procedure we used to quantize the primordial gravitational waves \cite{muka,giovannini,freitas} can be applied to the wave equation (\ref{eq:Phi}). This will transform the classical field $\Phi$ into the quantum field
\begin{equation}
   \hat{\Phi}=a^{\beta}\hat{\mu} \quad ,
\end{equation} 
which allow us, using the two-points correlation function $\left\langle \hat{\Phi}(\tau,\vec{x})\hat{\Phi}^{\dagger}(\tau,\vec{y})\right\rangle$, to find the quantum power spectrum
\begin{equation}
   \label{eq:powerPhihydro}
   P_{\Phi}(k,\tau)=\frac{k^3}{2 \pi^2}\frac{|\mu|^2}{a^{-2\beta}} \quad,
\end{equation}
with $\mu=\frac{\sqrt{\pi}}{2}\sqrt{|\tau|}\mathcal{H}_{\nu}^{(1)}(z)$.

\begin{figure}[ht]
   \centering
   \includegraphics[width=0.8\textwidth]{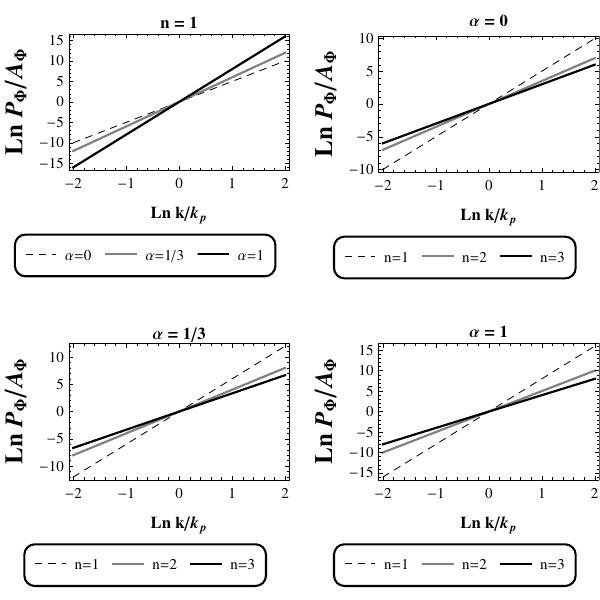}
   \caption{The power spectrum of the potential $\Phi$ for various values of the parameters $\alpha$ and $n$ as a function of the wavenumber $k$.}
   \label{fig:Phi3}
\end{figure}


We can also write the quantum power spectrum as
\begin{equation}
   P_{\Phi} \equiv A_{\Phi}\left(\frac{k}{k_p}\right)^{n_\Phi-1} \quad ,
\end{equation}
where $A_{\Phi}$ is the power spectrum amplitude, $k_p$ is some fixed pivot scale and $n_\Phi$ is spectral index, which can be calculated following the relation
\begin{equation}
   n_{\Phi}-1=\frac{d\ln{P_{\Phi}}}{d\ln{k}}\Bigg|_{aH=c_s k} \quad ,
\end{equation}
where $aH=c_s k$ is the horizon crossing condition, is given by
\begin{equation}
   \label{eq:indexPhihydro}
   n_{\Phi} = 1-2\left(\epsilon-\beta-1\right)\left(1-\epsilon\right)^{-1}\approx 3+2\beta\left(1+\epsilon\right) \approx -3c_s^2 \quad .
\end{equation}

In order to find the evolution of the density perturbation $\delta \rho$ we only need to perturb the conservation equations $T^{\mu\nu}_{~~;\nu}$ \cite{muka} to find the hydrodynamics perturbations
\begin{eqnarray}
   \label{eq:conspert1a}
   \delta \rho' +\mathbb{H}(\delta \rho +\delta p)-3\Phi'(\rho + p)+a(\rho+p)\delta u^{i}_{~,i} =0 \quad , \\
   \label{eq:conspert1b}
   a^{-4}\left[a^5(\rho +p)\delta u^{i}_{~,i}\right]'+\nabla^2\delta p+(\rho+p)\nabla^2\Phi=0 \quad .
\end{eqnarray}
In the regime $a \ll a_0$ the equation (\ref{eq:conspert1a}) will become
\begin{equation}
   \delta'+\left(1+c_s^2\right)\mathbb{H}\delta=0 \quad,
\end{equation}
where $\delta=\delta \rho/\rho$ is the density contrast, and the classical solution is
\begin{equation}
   \delta \propto a^{-\left(1+c_s^2\right)} \quad.
\end{equation}

To find the quantum evolution of the density contrast $\delta$ we need to find a wave equation. This can be done if we substitute the conservation equation (\ref{eq:conspert1a}) into (\ref{eq:conspert1b}) and consider the appropriate approximations to have
\begin{equation}
   \delta''+\left(5+c_s^2\right)\mathbb{H}\delta'+\left[k^2c_s^2-\left(1+c_s^2\right)\left(4\mathbb{H}^2+\mathbb{H}'\right)\right]\delta=0 \quad.
\end{equation}
Making the transformation $\delta = a^\gamma \mu$, where $\gamma=-\frac{1}{2}\left(5+c_s^2\right)=\frac{1}{2}\left(\frac{\alpha+1}{n}-4\right)$ we find the wave equation
\begin{equation}
   \tau^2\mu''+\left[\left(k\tau c_s\right)^2-\left(\gamma^2+9\gamma+20\right)\right]\mu=0 \quad,
\end{equation}
and finally, with the transformations
\begin{eqnarray}
  \mu =\sqrt{|\tau|}\mu \quad , \\ 
	z = k|\tau| c_s \quad,
\end{eqnarray}
we will find
\begin{equation}
   z^2\frac{d^2g}{dz^2}+z\frac{dg}{dz}+\left[z^2-\left(\gamma^2-9\gamma+81/4\right)\right]g=0 \quad ,
\end{equation}
which is the Bessel differential equation with the same solution (\ref{eq:solBessel}), with the order $\nu=\sqrt{\gamma^2-9\gamma+81/4}$.

We can use again the already discussed quantization process to make $\delta \rightarrow \hat{\delta}$ to find both power spectrum and spectral index
\begin{eqnarray}
   \label{eq:powerdelta}
   P_\delta(k,\tau) & = & \frac{k^3}{2\pi^2}\frac{|\mu|^2}{a^{-2\gamma}} \quad, \\ 
	 \label{eq:indexdelta}
	 n_\delta & = & 1-2\left(\epsilon-\gamma-1\right)\left(1-\epsilon\right)^{-1}\approx 3+2\gamma\left(1+\epsilon\right) \approx -(2+c_s^2) \quad ,
\end{eqnarray}
where $\mu$ is described in terms of the Hankel function as $\mu=\frac{\sqrt{\pi}}{2}\sqrt{|\tau|}\mathcal{H}_{\nu}^{(2)}(z)$, and 
\begin{equation}
   P_{\delta} \equiv A_{\delta}\left(\frac{k}{k_p}\right)^{n_\delta -1} \quad .
\end{equation}

\begin{figure}[ht]
   \centering
   \includegraphics[width=0.8\textwidth]{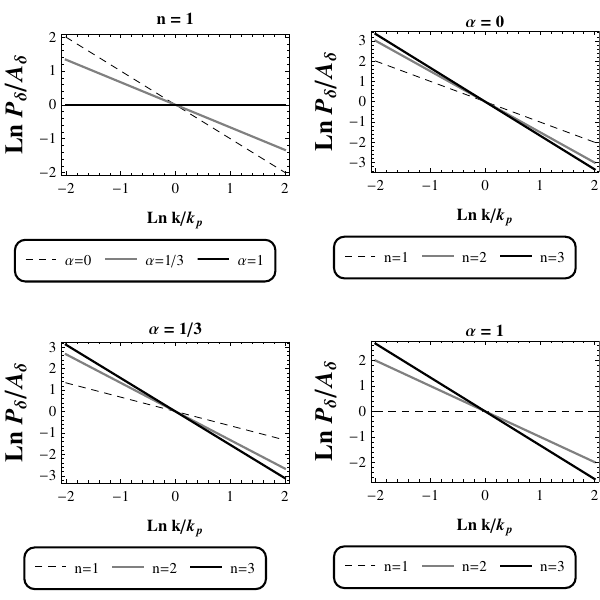}
   \caption{The power spectrum for the quantum density contrast for various parameters as a function of wavenumber.}
   \label{fig:contraste1}
\end{figure}


\section{Conclusions}                          
\label{sec:conclusao}                          

In this work we assumed that the primordial Universe was filled with a fluid described by the equation of state $p = \alpha\rho + k\rho^{1 + 1/n}$, (\ref{eq:EoS}), that is the sum of a standard linear equation of state and a polytropic term. The polytropic term, with $n = 1$, can be considered as a generalization of the standard Bose-Einstein condensate dark matter equation of state. Following \cite{chavanis2, chavanis3}, but letting the parameters $\alpha$ and $n$ free, we show that the EoS (\ref{eq:EoS}) can describe a inflationary Universe in the case of attractive self-interaction. We found the slow-roll parameters $\epsilon$ and $\eta$. 

In Figure \ref{fig:potencial} we plotted the scalar field's potential as a function of the scale factor ratio $a/a_0$. We can see that the difference between the panel with the curves representing dust ($\alpha = 0$) and radiation ($\alpha = 1/3$) are the smallest. The behavior follows a pattern where the case with $n = 1$ the concavity of the curve is downward while for $n = 2$ and $n = 3$ the concavity is upward. On the other hand, the panel with figures representing the behavior of rigid material, either for $n = 1$, $n = 2$ and $n = 3$ have the same concavity within the limits of the scale factor ratio $a/a_0$. Moreover, in this same figure we note that in a model of Universe filled only with dust and $n = 1$, which represents the BEC model, the slow-roll period ends earlier when compared with the stiff matter or radiation cases. As can be seen in Figure \ref{fig:potencial}, the case where $n=1$ the slow-roll period lasts longer than the cases with $n>1$. In all cases the inflationary era is longer for stiff matter ($\alpha=1$).

We calculated the slow-roll conditions for the scalar field during inflation (see Figure \ref{fig:epsilon}). We can see that in the model with stiff matter the slow-roll period, compared with the scale factor ratio $a/a_0$, is longer than the scenarios with dust $(\alpha = 0)$ and radiation $(\alpha = 1/3)$ for both $n = 1$ and $n = 2$. This situation indicates that the accelerated expansion of the Universe with stiff matter is slower than with dust and radiation. This characteristic may have important consequences in the process of evolution of the Universe, seen that the presence of stiff matter in FRW cosmological models produces an abundance of relic species of particles after the Big Bang due to the expansion and cooling of the Universe \cite{kami}. The presence of stiff matter in FRW cosmological models may also help explaining the baryon asymmetry and the density perturbations of the right amplitude for the large scale structure formation in our Universe \cite{joy}, and may also play a important role in the spectrum of relic gravity waves created during inflation \cite{sah}. These two important consequences may be changed due to the behavior of the slow-roll parameter $\epsilon$ for the model of stiff matter.

Figure \ref{fig:Phi3} shows the unnormalized power spectrum of $\Phi$ for various values of the parameters $\alpha$ and $n$ as a function of wavenumber $k$. We can clearly see that the curvature power spectrum is not scale invariant, and we see more power for small scales. It's spectral index (\ref{eq:indexPhihydro}) is proportional to the fluid's speed of sound, which, during inflation, can be approximated to
\begin{equation}
   c_s^2 \approx - \left(\frac{1+\alpha+n}{n}\right) \quad .
\end{equation}

Although we are dealing with an inflationary phase, which prevents agglomeration of matter, we also study the density contrast of the polytropic fluid, that has the quantum spectral index (which is also a function of the fluid's speed of sound) give by equation (\ref{eq:indexdelta}). The power spectrum behavior can be seen in Figure \ref{fig:contraste1}. In the left upper panel and in the right bottom panel of Figure \ref{fig:contraste1} we see that the power spectrum is scale invariant for $\alpha=1$ and $n=1$. For others combinations of $\alpha$ and $n$ we see more power for big scales.

It is known that when we consider a fluid with negative pressure, the equivalence between hydrodynamical and field representation exists only at the background level: at perturbative level, the model behave in a complete different way \cite{sergio}. Hence, in situations where negative pressures are concerned, a field representation leads to a much more complete scenario, being closer to a realistic model. We hope to present a more general analysis involving the comparison between the hydrodynamical model and the scalar representation of this polytropic equation of state in a future study, both in terms of background level and in the pertubative level in order to verify the behavior of the power spectrum in the early Universe.

To have a more robust analysis  of the behavior of the cosmological models with the polytropic equation of state we can use here the Bayesian chi-square $\chi^2$ minimization technique to limit the different parameters of the EoS for a viable cosmological model considering the observational data available today, but we have here a more complex situation: one primordial inflationary phase described by the polytropic equation of state with $k < 0$ and $n > 0$ and a current phase of accelerated expansion described by the same equation of state but with $k > 0$ and $n < 0$. In practice they are two different cosmological models. The idea was already developed in \cite{chavanis6, chavanis2, chavanis3}. We intend to use a combination of these two models in order to apply the statistical techniques mentioned above. The best-fit values of the model parameters are then determined from the chi-square function to study the evolution of the Universe. We plan to show this comparison with observations in a future work.

To summarize, the polytropic equation of state represents a interesting scenario to study the evolution of the Universe. The similarities with the models that are described by a linear equation of state, more than a simple coincidence, should be investigated with others kind of representations that not just the hydrodynamical representation and with statistical methods to verify the feasibility of the model.

\vspace{0.5cm}

\vspace{0.5cm} \noindent \textbf{Acknowledgments} \newline
\newline
\noindent
This work has received partial financial supporting from CNPq (Brazil) and CAPES (Brazil). We would like to express our gratitude to T. Battefeld and C.G.F. Silva for the helpful suggestions and for a careful reading of the manuscript. We are grateful to the
referee for his useful comments. R.C. Freitas thanks to the Institute for Astrophysics of the G\"ottingen University (Germany) for the kind hospitality during the time this paper was completed.

\end{document}